**Rolled-up self-assembly of compact magnetic inductors, transformers and resonators.**

*Dmitriy D. Karnaushenko[#], Daniil Karnaushenko[#,]*, Hans-Joachim Grafe, Vladislav Kataev, Bernd Büchner, Oliver G. Schmidt*

Dr. D. D. Karnaushenko, Dr. D. Karnaushenko, Prof. Dr. O. G. Schmidt
Institute for Integrative Nanosciences, Institute for Solid State and Materials Research Dresden (IFW Dresden), 01069 Dresden, Germany
E-Mail: d.karnaushenko@ifw-dresden.de

Dr. H.-J. Grafe, Dr. V. Kataev, Prof. Dr. B. Büchner
Institute for Solid state Research, Institute for Solid State and Materials Research Dresden (IFW Dresden), 01069 Dresden, Germany

Prof. Dr. O. G. Schmidt
Material Systems for Nanoelectronics, Chemnitz University of Technology, 09107 Chemnitz, Germany
Center for Advancing Electronics Dresden, Dresden University of Technology, 01062 Dresden, Germany

[#]Authors contributed equally to this work.
*Corresponding author.

*Three-dimensional self-assembly of lithographically patterned ultrathin films opens a path to manufacture microelectronic architectures with functionalities and integration schemes not accessible by conventional two-dimensional technologies. Among other microelectronic components, inductances, transformers, antennas and resonators often rely on three-dimensional configurations and interactions with electromagnetic fields requiring exponential fabrication efforts when downscaled to the micrometer range. Here, the controlled self-assembly of functional structures is demonstrated. By rolling-up ultrathin films into cylindrically shaped microelectronic devices we realized electromagnetic resonators, inductive and mutually coupled coils. Electrical performance of these devices is improved purely by transformation of a planar into a cylindrical geometry. This is accompanied by an overall downscaling of the device footprint area by more than 50 times. Application of compact self-assembled microstructures has significant impact on electronics, reducing size, fabrication efforts, and offering a wealth of new features in devices by 3D shaping.*

Strain-driven self-assembly of thin films into three dimensional geometries has become a powerful technique to produce a plethora of differently shaped microarchitectures.[1,2] In combination with established micropatterning techniques, highly parallel processing of ultra-compact three-dimensional devices on a single chip has been demonstrated.[3,4] Potential application areas of such mass-produced components are manifold ranging from basic microelectronic components and sensors to systems and robotics both on and off the chip.[1,5–7]

A special and particularly interesting category of micro-origami is the strain-driven roll-up of patterned layer stacks into microtubular devices.[8,9] Such devices have been employed for fabrication of capacitors[10], inductors[11], transistors[12], sensors[13], and microbots[14], to only name a few. Rolled-up nanotechnology has recently seen significant progress driven by cleverly designed layer stacks and material choices. The introduction of a polymer based platform in combination with advanced functional materials and device layouts has led to fully integrated biomimetic circuitry[15], microtubular GMI sensors[16] and impedance matched antenna arrays.[17] However, the complexities associated with the self-assembly of microorigami devices and overall process compatibility remains challenging requiring the right choice of materials and establishing predictable and viable design rules. For instance, reproducibility of the geometry is one of the key challenges, which affects functionality of a final device and must be precisely controlled in order to achieve high yield and reliability of electronic systems.[15,17]

Here, we demonstrate advanced roll-up assemblies for a new category of devices. We create inductive and inductively coupled systems for applications in transformers and ultra-compact high quality electromagnetic resonators for nuclear magnetic (NMR) and electron spin (ESR) resonance characterization set-ups. Normally, fabrication of these devices require numerous fabrication steps[18,19] in a conventional planar microelectronic process.[20,21] In contrast, application of the parallel self-assembly strategy reduces fabrication efforts and footprint area[11], and enhances performance up to an order in magnitude if compared to their planar counterparts. The geometry of the assembled rolled-up stack shown in **Fig 1 a** contains polymeric and metallic parts that finally reveal magnetic inductive coils. The initial planar stack consists of four patterned, functional, shapeable, ultrathin, polymeric layers with a rectangular geometry having length (L) and width (W) of 17 mm and 5.5 mm respectively. The layers contain the sacrificial layer (SL), hydrogel layer (HG) and two rigid reinforcing layers made out of polyimide (PI) material (**Fig. 1 b**). Each of the two rigid layers was patterned using a specially designed structure required for a correct and reproducible self-assembly of the overall architecture into a cylinder. Normally, structures with such an aspect ratio and geometry do not assemble properly into tubular shapes, rather fold from sides and transform into polygonal geometries.[22,23] To avoid such problems a number of bracket structures and crack propagating edges (CPEs) were implemented on both sides of the rigid layer in order to guide the self-assembly process (**Fig 1. b**) in a desired direction during the batch wafer (100 x 100 x 1 mm$^3$) fabrication process. The optimal spacing of ~200 µm between every couple of brackets was determined considering the desired final circumference of the microtube.[17,22,23] The CPEs, introduced in the double reinforcing layer (PI 1 and 2 with 1:1 thickness ratio), ensures a controlled rupture of the bracket structures during the rolling process.

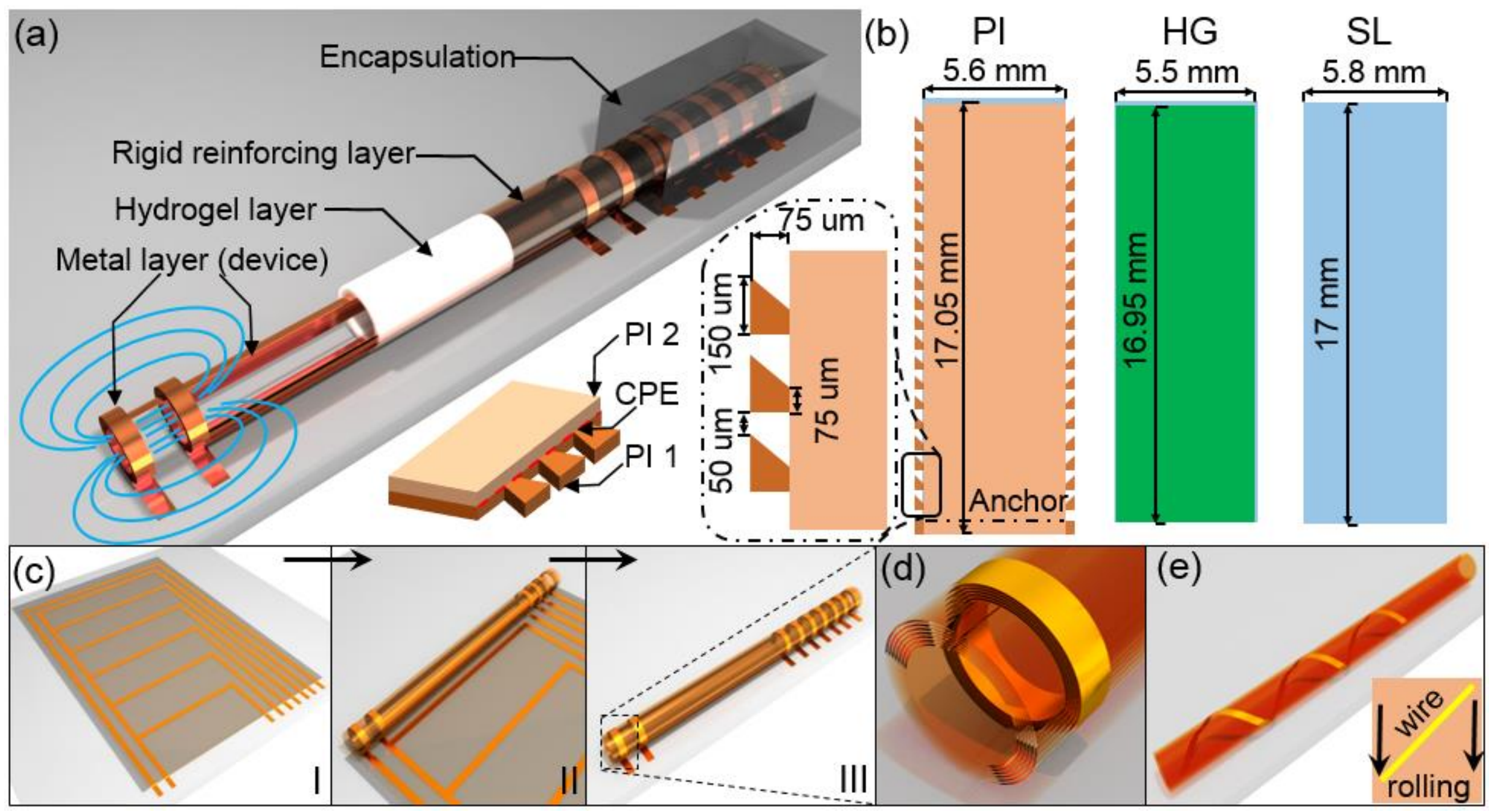

**Figure 1.** Illustration of the self-assembly of functional 3D magnetic components using shapeable ultrathin films. (**a**) Self-assembled and encapsulated 3D device revealing coils and their mutually coupled configurations. (**b**) Shapeable layer stack consisting of three distinct polymeric structures, namely the sacrificial layer (SL), the hydrogel layer (HG) and the reinforcing polyimide layer (PI). The latter possesses a two layer structure where the first layer (PI 1) is equipped with fixators and the second (PI 2) layer develops crack propagation edges (CPEs), which helps to release the brackets during the rolling process and guide the overall (**c**) self-assembly process. (**d**) Rolling of a straight wire leads to a Swiss Roll coil, (**e**) while rolling of the tilted wire forms a helix geometry.

We applied different planar layouts of conductors (see e.g. **Fig. 1 c I**) on the surface of these layered shapeable materials. The roll-up self-assembly leads to a Swiss-roll architecture (Fig. 1 c I-III, d) with a final diameter of ~300 µm, winding thickness of 4 µm and a roll width W of 5.5 mm. Upon self-assembly the footprint of the structure is effectively reduced from about 100 mm$^2$ to 1.5 mm$^2$ (**Fig 1. c**). The inductance of a conductor in the assembled state depends on its orientation on the planar surface (compare **Fig. 1 c, d and e**) and is maximum when the device is assembled into a zero pitch coil (**Fig.1 d**). The planar design of the conductor should contain, due to the constraints imposed by the Swiss-roll geometry, a couple of straight conductors with a perpendicular section that electrically connect both coils after the self-assembly. This so called "**Π**" geometry allows (see **Fig. 1 c**) electrical feeding from the "anchor" side of the device attached to the rigid support. We have prepared a number of such shapes (**Fig. 2 a-e**) made out of Ti – Cu – Ti with a conductor cross section of 100 x 3 $\mu m^2$ in a single lithographic run (the exact geometry is shown in **Sup. Fig. 1**) and self-assembled into Swiss-rolls in a batch wafer scale process (**Fig. 2 b**). During this process, the planar "**Π**" shaped conductors were transformed into a number of coils corresponding to the equivalent circuit shown in **Fig. 2 c**. Each conductor, oriented along the rolling direction, forms a coil, while the perpendicular conductors do not reshape during this process and just provide an electrical connection between the left (**Fig. 2 d group 1**) and right **(Fig. 2 e group 2**) set of coils (**Fig. 1 a, 2 a**). Each conductor possesses a DC resistance ranging from 1 to 10 Ohm in both the planar and 3D assembled states corresponding to the cross section and geometry of the conductors. With this resistance value it was possible to push up to 300 mA of AC current without any thermal damage of the 3D structures. The value of the quality factor of the inductors ranges from 0.5 to 5 demonstrating an average enhancement in the performance of up to 2.5 times due to the rolling process, measured at 10 MHz (the spectrum is shown in **Sup. Fig. 2**). It is clear from the geometry of the planar layout that the self and mutual inductances of each planar loop should have a direct relation to the wire length and the area surrounded by the wire. For instance, the planar conductor patch between electrodes (0) and (8) has the initial value of the planar inductance equivalent to $L_{PL(0)-(8)}$ = 35.5 nH at 10 MHz. The loop between the electrodes (3) and (7) possesses an inductance $L_{PL(3)-(7)}$ = 36.5 nH – a little bit larger than $L_{PL(0)-(8)}$ due to the longer length of the conducting path. The measured inductance $L_{PL(1)-(2)}$ of the smallest loop between the electrodes (1) and (2) has a value of only 11 nH.

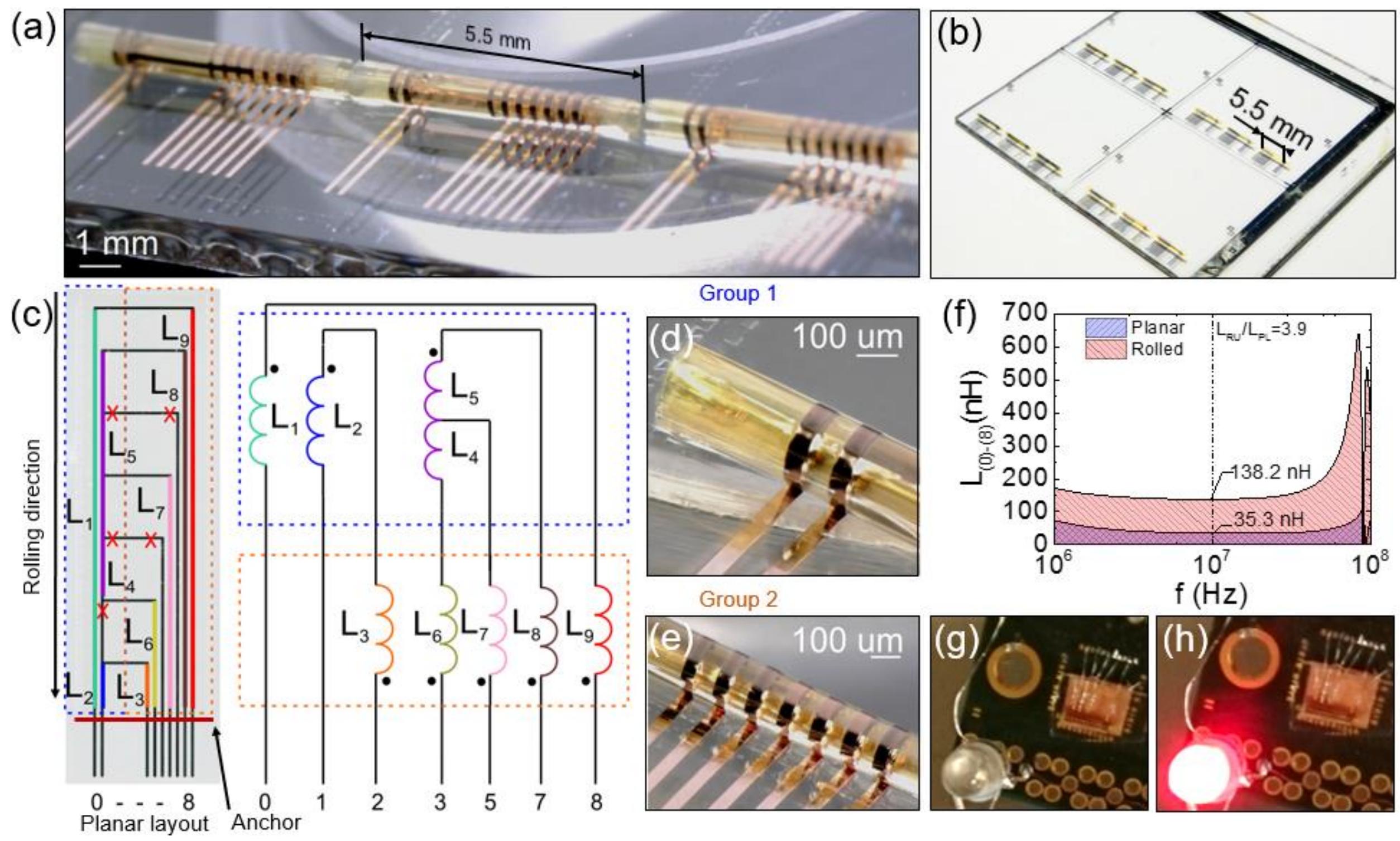

**Figure 2.** Self-assembled micro coil structures that include inductors and their mutually coupled configurations – transformers. (**a**) Three equivalent rolled up structures fabricated in a single run on the glass substrates using (**b**) batch fabrication and self-assembly process. (**c**) Planar layout of conductors forming a set of coils and its representative equivalent circuit. The circuit consists of (**d**) left and (**e**) right side of coil groups. (**f**) Inductance spectra demonstrating substantial (3.9 times) enhancement of the self-inductance between the electrodes (0) and (8) due to the self-assembly process.

The self-assembly into the tubular structure (**Fig. 2 c**) converts the overall inductance of the wire $L_{PL(0)-(8)}$ into the sum of the coil inductances $L_{RU(0)-(8)} = L_1 + L_9 = 138.2$ nH that includes the inductance of the straight section as well. As can be recognized from the obtained characteristics (**Fig. 2 f**), the inductance of $L_{RU(0)-(8)}$ in the assembled state is 3.9 times higher if compared to the $L_{PL(0)-(8)}$ inductance of its initially planar counterpart. Similarly, two other inductances in the rolled state demonstrate increased values of their inductances, ~2.5 times for $L_{RU(3)-(7)} = L_6 + L_4 + L_5 + L_8 = 91.4$ nH and ~3.3 times for $L_{RU(1)-(2)} = L_2 + L_3 = 35.9$ nH compared to their planar counterparts respectively at 10 MHz (the spectra are shown in **Sup. Fig. 3**). Here, the inductance ratio between the rolled and planar geometries is affected by the series inductance of the straight conductor (parallel to the Swiss-roll axis) according to $\frac{L_{RU}+L_C}{L_{PL}+L_C}$, where $L_C$, $L_{PL}$ and $L_{RU}$ are the inductances of the straight conductor, the same conductor in the planar and rolled states respectively. The inductance of the straight conductor is obviously not altered due to the rolling process, thus strongly affecting the inductance ratio of the structure between the electrodes (3) and (7), but not the structure between the electrodes (1) and (2) (see **Fig. 2 c**). The ratio between the inductances in the rolled and planar states provides a good merit for the self-assembly process that has to be carefully analyzed during the design for the sake of maximum final performance of 3D assembled devices. Besides, this analysis reveals an optimal 2D layout of conductors allowing to effectively exploit the available 3D space. In this respect, the self-assembly of the coils demonstrates substantial enhancement of the inductance due to the geometric transformation and shrinking of the overall footprint area by up to 60 times.

The rolled coils (**Fig. 1 a, 2 a, c - e**) e.g. $L_1$ & $L_2$, $L_1$ & $L_4$ & $L_5$ and $L_8$ & $L_9$ are mutually coupled, due to their close proximity. This is obvious from the circuit shown in **Fig. 2 c.** The coil set ($L_1$ & $L_9$) between the electrodes (0) and (8) shares a mutual inductance with the coil set ($L_5 + L_4$ & $L_8$) formed between the electrodes (3) and (7) respectively $M_{RU(08)-(37)}$, which is equivalent to 18.8 nH. The mutual inductance of the rolled structure is higher, compared to the mutual inductance $M_{PL(08)-(37)} = 12$ nH of the initially planar structure. A similar character is observed for the set of coils between the electrodes (0) and (8) coupled to the set of coils between the electrodes (1) and (2), revealing more than twice an enhancement of the value of the mutual inductance in the rolled state $M_{RU(12)-(08)} = 3.9$ nH compared to $M_{PL(12)-(08)} = 1.5$ nH of the planar structure (the spectra are shown in **Sup. Fig. 4 a-d**). The mutual inductances of the planar and rolled geometries have, however, different origins. The flux linkage in the planar loop structure is mostly perpendicular to the surface, which vanishes in the self-assembled structure. Due to the rolling process, the set of axial coils is formed with axially oriented flux, allowing inductive coupling of these coils. This aspect is appealing for fabrication of axial micro transformer structures in the self-assembled state (**Fig. 2 c**). The key parameter of these devices is the coupling coefficient that shows how much of the total flux is shared among the coils. A coupling coefficient $k_{RU(08)-(37)} = 16.8\%$ was measured for the same set of coils, which in the rolled state of the device is half of the coupling coefficient $k_{PL(08)-(37)} = \sim 33.4\%$ of the initially planar structure (the spectra are shown in **Sup. Fig. 5 a-d**). However, an opposite behavior is observed for the set of coils between electrodes (0) and (8), and the set of coils between electrodes (1) and (2). This set of mutually coupled coils demonstrates enhancement of the coupling coefficient in the rolled state $k_{RU(08)-(12)}=9.8\%$ (spectra are shown in **Sup. Fig. 5 b**), over the low coupling coefficient of only $k_{PL(08)-(12)} = 7.6\%$ in the planar state. Such a difference in the behavior of the two coupled sets of coils can be understood from the layout of their planar conductors (**Fig. 2 c**) and the definition of the coupling coefficient, which is $k=M/\sqrt{L_P L_S}$, where M, $L_P$ and $L_S$ are the mutual and self-inductances of the primary and secondary coil sets respectively. An increase in the self-inductance, as it occurs during roll-up, should be accompanied by an increase of the mutual inductance in order to keep the coupling coefficient constant. This is not the case for the partial inductance $L_6$ (**Fig. 2 c**) located further away from the partial inductance $L_1$, compared to the positions of the partial inductances $L_4$ and $L_5$. Thus, the increase of the inductance $L_6$ cannot be compensated during the rolling by its vanishing mutual inductance with the inductance $L_1$, therefore $k_{RU(08)-(37)} < k_{PL(08)-(37)}$. The axial proximity of coil $L_6$ to $L_4$ and $L_5$ also explains the only 2.5 times enhancement of the $L_{RU(3)-(7)}$ self-inductance compared to other coil sets. In this respect, coil $L_6$ has a vanishing mutual inductance with the inductances $L_4$ and $L_5$, negatively affecting the overall inductance increase between electrodes (3) and (7). This difference clearly demonstrates the demand of accurate design rules for the planar layout by carefully defining the position and orientation of the conductor structures directly affecting the final characteristics of the self-assembled 3D device.

For many applications, including signaling circuits as well as power converters, a particularly interesting configuration of mutually coupled coils is the centered-tap-secondary transformer.[24] This configuration mainly relies on inductances $L_1$, $L_4$ and $L_5$ in the self-assembled device (**Fig. 2 c**). The windings formed by the inductances $L_6$, $L_7$ and $L_8$ should have much less coupling due to the twice larger distance between them. Otherwise, inductances $L_9$ and $L_8$ should have a strong mutual coupling that contributes to the overall coupling between the conductors (5) - (7) and (0) - (8). The measured characteristics confirm this behavior (the spectra are shown in **Sup. Fig. 3 c, d** and **4 c, d**). The self-inductance value and the inductance ratio of the set between (3) - (5) is much smaller ($L_{RU(3)-(5)}$ = 37.2 nH and $L_{RU}/L_{PL}$ =1.9 respectively) compared to the set between (5) - (7) ($L_{RU(5)-(7)}$ = 83.7 nH and $L_{RU}/L_{PL}$ =2.8 respectively), which is purely due to the geometry of the initially planar conductor containing several small sections with vanishing self- and mutual coupling among them in the self-assembled state. Sections (3) - (5) and (5) - (7) differ by almost a factor of two in their coupling coefficient $k_{RU(35)-(08)}$=13.5% and $k_{RU(57)-(08)}$=25.6% with the section (0) – (8), which is obvious considering the relative positions (**Fig. 2 c**) of their conductors (the spectra are shown in **Sup. Fig. 5 c, d**). Thus, it is very critical to consider the right design of the 2D layout in order to achieve symmetric operation of such an electronic device.

Starting at about 40 MHz, the mutually coupled inductors reveal an increase and then drop in their inductances to almost 0 H at a frequency of about 97 MHz (the spectra shown in **Fig. 2 f** and spectra's shown in **Sup. Fig. 3**). This happens due to the resonance condition promoted by the intrinsic capacitance of the rolled-up coils. For demonstration purposes we energized an LED connected to the secondary side

(0) - (8) of the micro transformer by powering the primary (3) - (7) side (detailed circuit in **Sup. Fig. 6**). The resonance frequency of the primary and secondary coil sets was matched to the frequency of ~40 - 50 MHz with external capacitors of about 100 pF connected in parallel on either side of the transformer. At the resonance frequency the energy can be transferred with higher efficiency[25] thus the LED can generate light (**Fig. 2 g, h**) rectifying the power transferred over the mutual flux between the self-assembled micro coils. The overall insulation among windings was measured to be more than 4 MOhm at 32 V, which was characterized by the source measuring unit Agilent B2902. All the impedance characterizations were performed by the two port network analyzer Agilent ENA5071 with an appropriate calibration routine.

We have applied the micro scale self-assembled coil as an NMR transducer to probe nuclear spin states in a small volume of a model material. The self-assembled micro coil was introduced in a commercial NMR probe forming an LC resonator. The resonance frequency of the resonator was adjusted with an external set of capacitors to the Larmor frequency of 1H nuclear spins processing in an external magnetic field. In this configuration the micro coil is able to transfer energy into the material subsystem and receive the response signal. For demonstration purposes we have integrated the tuned transducer (**Fig. 3 a, b**) into a commercial NMR setup equipped with 3 T and 7 T superconducting magnets. We used glycerin in the inner opening of the rolled up structure (**Fig. 3 a**) for the test measurement. Glycerin shows two hydrogen (1H) peaks at about 3 ppm and 4.5 ppm, which are revealed by a standard solenoid copper coil possessing a diameter of d = 0.6 mm and a length of l = 2 mm. (**Fig. 3 c**). In the same figure we demonstrate the first successful NMR measurements performed with the self-assembled micro coil. The signal shows a clear signature of 1H in glycerin at the correct resonance frequency. The broader line width in the obtained signal for the rolled up micro coil compared to the standard solenoid copper coil is expected for small scale coils and just confirms the functionality of the transducer. Possessing a strong field strength such coils are affected by different local magnetic susceptibilities of the coil itself and of the materials surrounding the coil. This effect is well known[26] and increases with the downscaling of the coil size where the bulk of the sample more closely approaches the coil windings[27].

Finally, we demonstrate the feasibility to realize high quality LC resonators without using external capacitors just relying on an alternative planar layout of the conductor to form a parallel plate capacitor and the coil in a single batch self-assembly process (**Fig. 3 d**). The resonator, in its 3D shape, was designed using an FEM model based on the ANSYS Academic software package (**Fig. 3 e**) and then fabricated in the same process as the micro coils with the only difference in the planar layout of the conductor. The most prominent design (**Fig. 3 e, f**) achieved a quality factor >40000 (**Fig. 3 g**). In this design, the coil stripe (**Fig. 3 e** right and the planar layout is shown on **Sup. Fig. 1 b**) is shifted away from the capacitor stripe (**Fig. 3 e** left) in order to ensure spatial separation of electric and magnetic fields in the assembled state that avoids radiation of electromagnetic waves. The self-assembled structure contains a cylindrical parallel plate capacitor (**Fig. 3 e**) and the multiwinding inductor, resulting in a high quality factor resonator tank.

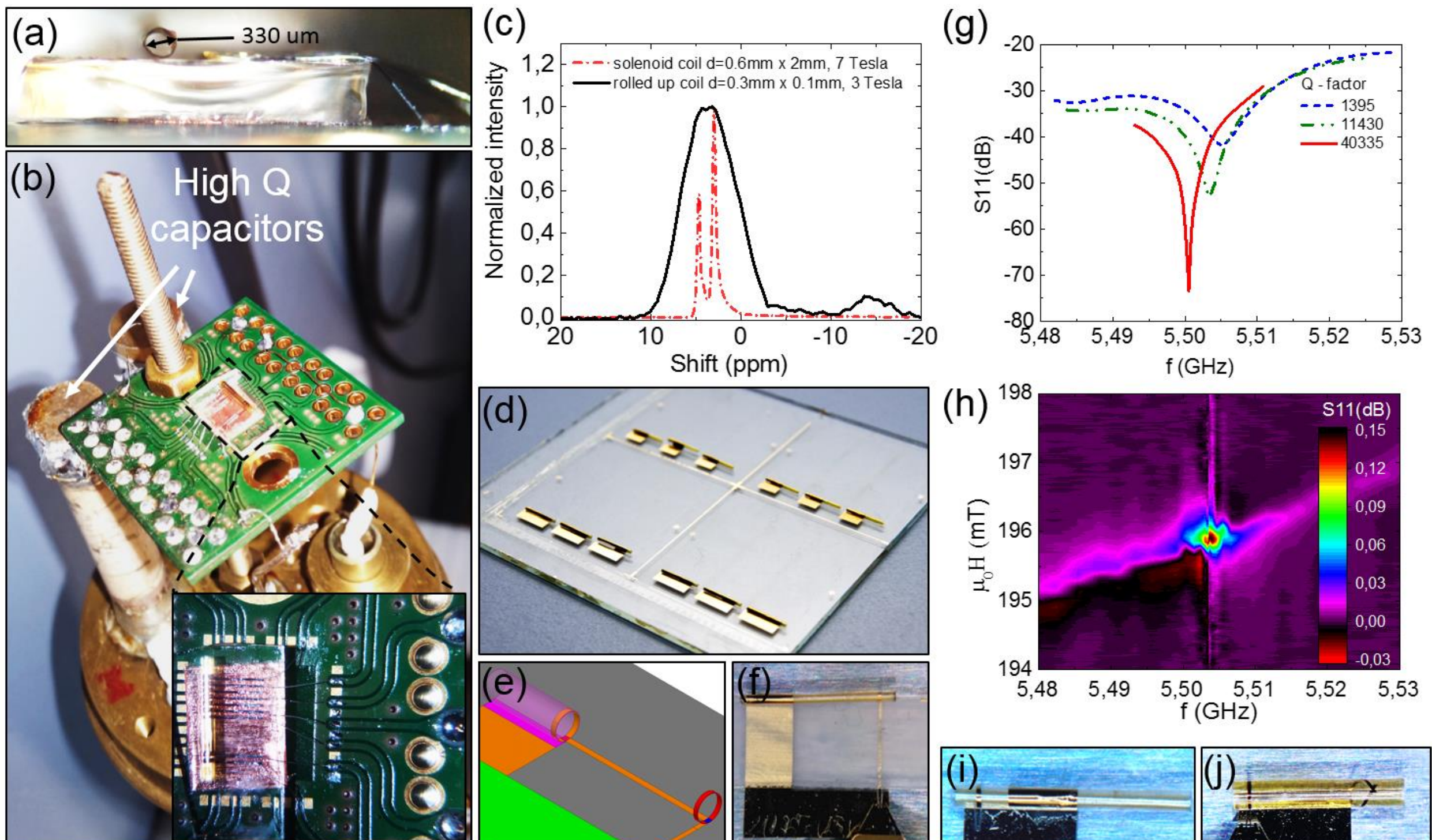

**Figure 3.** Self-assembled high quality micro coils and micro resonators applied for NMR and ESR characterization. (**a**) Micro tubular self-assembled architecture possessing an opening of 330 μm suitable for insertion of a small amount of an NMR sample. (**b**) Rolled up self-assembled devices integrated into a commercial NMR probe and connected in a circuit with high-Q capacitors forming an LC resonator. (inset) Magnified view of the device demonstrating the soldered connections directly to the printed circuit board. (**c**) NMR response of the rolled coil and sampled with a glycerol shows characteristic peaks at the correct positions, which is compared to a regular Cu wire solenoid. (**d**) Self-assembled high quality micro resonators that were fabricated in a batch wafer scale process. (**e**) Optimal design of the 3D resonator was simulated and (**f**) fabricated using shapeable ultrathin films. (**g**) The high-Q resonator demonstrates variation in the resonance frequency and the quality factor among three fabricated in a single run devices. This happen due to the slight variations in the geometry of the devices. The highest Q resonator was applied to measure the ESR spectrum. (**h**) Measured response peak of the DPPH radicals in magnetic field around 196 mT and f = ~5.5 GHz. Several versions of rolled up structures (**i, j**) revealing various geometries of conductors.

For testing, a simple ESR setup was built integrating the micro coil ESR resonator (**Fig. 3 f**). The setup was equipped with a strong biasing electromagnet, a shimming couple of coils and small field scanning coils supporting a very precise settling of the magnetic field between the pole shoes (±2.5 μT). An additional filter circuit and a custom designed nonmagnetic probe were implemented in order to eliminate residual magnetic inhomogeneity's and noise coming from the power supply of the magnet, which otherwise disturbs characterization of the high Q resonances presented in the ESR model sample. We used 2,2-diphenyl-1-picrylhydrazyl (DPPH) as the model material, which is an important functional agent in chemistry i.e. for ESR monitoring of the antioxidation activity of biologically relevant substances.[28] The material was loaded into the resonator and characterized with an applied magnetic field, which was swept around 196 mT. The central field of 196 mT corresponds to the ESR resonance of DPPH free radicals at about 5.5 GHz which is close to the resonance peak of the chosen resonator (**Fig. 3 h**). We had to choose the field and an appropriate frequency range between 5.48 GHz and 5.53 GHz measuring return loss (S11) to find resonance characteristics of the micro resonators (**Fig. 3 g**). The variation in characteristics of the resonators accounts for a slight sensitivity of the high Q device to a deviation of geometric parameters of the rolled up structure. A subtle variation in the diameter or the winding misalignment can be dramatic for the device performance affecting the capacitance or the inductance.[17] The resonator design is sufficiently tolerant towards this issue (**Fig. 1 a**) purely relying on the planar conductor design among other less stable structures (**Fig. 3 i, j**).

In conclusion, we have designed and fabricated rolled-up cylindrical micro coils, transformers and resonators relying on shapeable polymeric ultrathin films. As the result we achieved up to 48 different devices on a single square (100 x 100 $mm^2$) shaped wafer. We could demonstrate a fully parallel wafer scale process utilizing the same fabrication routine, shapeable layer stack and the 3D tubular geometry. The self-assembly of very simple conductor structures into the Swiss-roll geometry provided means to fabricate coils, transformers and high quality resonators. The overall process allows omitting a number of intermediate steps, which are otherwise required in conventional 2D processing schemes. We showed for the first time that shaping of the initially planar structure like a conductor can lead to >50 times more compact 3D inductive coils with enhanced up to 4 times inductance, reaching >100 nH at MHz regime. The mutual inductance among some inductors and the coupling coefficient were enhanced promoting applications such as ultra-compact transformers. We envision applications of these magnetic self-assembled devices in microelectronics, radio frequency, communication devices and power converters. Fully integrated systems will employ micro scale inductances, transformers, and resonators operating as e.g. impedance matching networks, filters or -- if combined with active electronics[15] -- fully integrated compact DC-DC converters. Additionally, sensitive micro coil electromagnetic resonators were realized, and their application for micro scale NMR and ESR spectroscopies was demonstrated. Application of NMR inert materials and other measures to create a homogeneous magnetic surrounding of the micro coil will substantially increase the resolution of the rolled-up NMR micro coils. Shimming and gradient coils can be accommodated within the same geometry to define the magnetic field profile of the resonators. These devices are promising if combined with sampling systems such as microfluidics or micro catheters for realization of compact resonance spectroscopies and imaging techniques of micro scale samples.

**Experimental Section**

*(1) Treatment of substrates:*

Square shaped glasses of 100 x 100 x 1 $mm^3$ were used as the substrates in this work (D263T eco glass, SCHOTT AG, Mainz, Germany). Initially, all the substrates were washed in the professional washer DS 500 (STEELCO S.p.A., Riese Pio, Italy) to remove all of the organic and inorganic contaminants, presented in a form of dust or films. Then, the surface was activated with oxygen plasma in the GIGAbatch 310M (PVA Metrology & Plasma Solutions GmbH, Wettenberg, Germany). This ensures further chemical surface modification with a monolayers of 3-(trimethoxysilyl) propyl methacrylate (TMSPM). For this, the glasses were placed in the vacuum oven at 150°C for 2h together with 150 µL of TMSPM.

*(2) Sacrificial layer:*

The sacrificial layer was prepared from acrilic acid (AA) and hydrated $LaCl_3$ obtained from Alfa Aesar UK and used without further purification. Firstly, 10 g of AA was partially neutralised using sodium hydroxide (Sigma-Aldrich Co. LLC., Germany) until the solution reaches pH = 5.5. The pH value of the solution was monitored using the pH meter CyberScan PC510 (Eutech Instruments Pte Ltd., Singapore). Than we added 7.36 g of $LaCl_3 \times 7\ H_2O$ to this solution, dissolved in 50 ml of DI to achieve the proportion of AA : La = 7:1. The AA : La solution was stirred for 30 min using a magnetic stirrer and during this time pH was slowly increased to 10 by dropping of NaOH. At elevated pH lanthanum complex precipitates and collected using a filter paper and a vacuum exicator with subsequent washing in DI water until litmus test paper (DuoTest, Macherey-Nagel GmbH & Co. KG, Germany) shows neutral reaction. White precipitate was collected and dried under nitrogen atmosphere resulting in a white powder of 4.68 g. As a final step, the dried and milled material was dissolved in AA at the concentration of 33% (wt/wt), photosensitized using 2% (wt/wt) of 2-Benzyl-2-(dimethylamino)-4-morpholinobutyrophenone and 4% (wt/wt) methyl diethanolamine (MDEA) (Sigma-Aldrich Co. LLC., Germany). Achieved solution was applied on the substrate using spincoating at 3000 rpm for 35s with the acceleration of 500 rpm/s. After predrying at 35°C for 5 min it was exposed with a broadband UV mask aligner MJB4 (SÜSS MicroTec SE, Garching, Germany) for 30 sec and then developed in DI water for 30 sec. In order to remove water insoluble rests of photo initiator a subsequent washing in propylene glycol monomethyl ether acetate (PGMEA) (Sigma-Aldrich Co. LLC., Germany) was done for 5 sec. Finally, samples were hard baked at 220°C for 5 min revealing high resolution structures of the sacrificial layer on the glass surface.

*(3) Shapeable bilayers:*

The assembly of the planer stack occurs in the bi-layer polymeric system due the swelling of the hydrogel layer (HG) reinforced by the polyimide layer (PI) in a basic conditions pH= 9 of the rolling solution. Preparation of the bilayer started from the synthesis of HG using poly(ethylene-alt-maleic anhydride) (PEMA) and N-(2-hydroxyethyl)methacrylate (HEMA) (Sigma-Aldrich Co. LLC., Germany). At first 6 g of PEMA has been dissolved in the 50 ml of N,N-Dimethylacetamide (DMAc) (Sigma-Aldrich Co. LLC., Germany), then to the solution 5.66 g of HEMA was added. All components were mixed thoroughly for 24 h at room temperature by roll-mixer to finish the reaction.

The PI was prepared by reaction of 12 g 4,4'-Methylenedianiline (MDA) (Sigma-Aldrich Co. LLC., Germany) and 19.65 g of 3,3',4,4'-Benzophenonetetracarboxylic dianhydride (BPDA) (Sigma-Aldrich Co. LLC., Germany). Both monomers were dissolved in 20 ml DMAc and gradually over an hour reacted together under culling conditions at 15°C, resulting in the highly viscous solution. Then a dimethylaminoethyl methacrylate (DMAEMA) (Sigma-Aldrich Co. LLC., Germany) added in a 1:1 molar ratio to the number of carboxylic groups in BPDA. The solution was left in the mixer for 4 h under cooling conditions to complete the reaction. At the end HG and PI solutions were photosensitized with 4% (wt/wt) of 2-Benzyl-2-(dimethylamino)-4-morpholinobutyrophenone (Sigma-Aldrich Co. LLC., Germany) and additionally diluted to have thickness in the range of 500 nm and 1500 nm respectively at 3000 rpm.

The bilayer stack was formed in a sequential photoparnening process of HG and PI. Materials in amount of 3 ml were applied on the substrate through 1 µm filter using following settings: polymer pre-spinning step was made at 500 rpm with acceleration 50 rpm/s for 10 s, main spinning is done at 3000 rpm for 30 s with acceleration of 500 rpm/s. The final post spinning step was done with acceleration of 2500 rpm/s for 3 s reaching 5000 rpm minimizing edge-beads. Then, polymers were prebacked at 40 °C for 5 min and 50 °C for 10 min respectively. After the drying step substrates were exposed using MJB4 mask aligner for 1.5 min and then developed. For HG layer development was done in diethylene glycol methyl ether (DEGMEE) (Sigma-Aldrich Chemie GmbH, Munich, Germany) for 60 sec with washing in PGMEA for 5 sec. PI layers were developed in the solution of solvents: 40 ml of NEP (Sigma-Aldrich Chemie GmbH), 20 ml of DEGMEE and 10 ml of ethanol (VWR International GmbH) for 90 sec with a subsequent washing in PGMEA for 5 sec. After the development layers were hard baked at 220°C for 10min.

*(4) Metal layers:*

The metal layer stack was patterned via etching process. The layer stack consisting of $Ti^{10\ nm}/Cu^{100\ nm}/Ti^{10\ nm}$ was deposited using magnetron sputtering at room temperature. For sputtering, Ar gas at partial pressure of $10^{-3}$ mbar and the base pressure $2.3x10^{-6}$ mbar was used. Then layer of 1 um AZ5214E photoresist (Microchemicals GmbH, Ulm, Germany) was patterned following the protocol provided by the manufacturer. Titanium etching process was made in the solution of sodium fluoride (NaF), ammonium peroxydisulfate and DI water in the proportion of 1 : 1 (mol/L) in 200 mL of water. Copper etching process was done during 5 s in the solution of HCl : $H_2O_2$ : $H_2O$ in a proportion of 1:2:10 (vol./ vol./ vol.).

*(5) Self-assembly rolled-up process of tubular architectures:*

The 2D layouts of polymeric structures with electronics layer were self-assembled into 3D Swiss-rolls by a selective etching of the SL in the solution of a strong chelating agent containing 15 g of sodium diethylenetriaminepenta-acetic acid (DETPA) (Alfa Aesar, UK) in 500 ml of DI water and a subsequent swelling of the HG. For dissolution of the chelate in water it was neutralized with sodium hydroxide in an amount sufficient to reach pH = 5.5. Additionally 10% of benzotriazole (Sigma-Aldrich Co. LLC., Germany) was introduced into the solution in order to inhibit etching of Cu layer. After the SL etching process rolling was performed in an equivalent DETPA solution, but having pH = 8.

*(6) Electrical characterization:*

All prepared self-assembled structures were characterized acquiring theirs S and Z parameters using the Cascade PM-8 probe station and vector network analyzer Agilent ENA 5071 (Agilent Technologies GmbH & Co.KG, Waldbronn, Germany) in the frequency range from 300 kHz to 100 MHz All the electrical connections with rolled-up strucures were made using ground signal ground (GSG) CASCADE |Z|-probes. The LED experiment was performed using a signal generator MG3692B (Anritsu Corporation, Kanagawa, Japan). To cancel any adverse effect of the interaction probe system with the magnetic field, the ESR measurements were done in an in-house made magnet, the probe and probe holder were made of non-magnetic materials. The electromagnet with a magnetic field in the range from 0 to 600 mT and maximum deviation less than 2.5 μT was achieved. The magnetic field was swept using an additional set of coils with the Keithley source meter (Tektronix UK Ltd., UK). Magnetic field was measured with the MAGSYS hall sensor (MAGSYS magnet systeme GmbH, Dortmund, Germany), where the probe sensor was closely placed to the sample in order to eliminate any differences in the readings of magnetic field and estimation of the Larmor precession frequency.

## Acknowledgements

We thank C. Krien and I. Fiering (IFW Dresden) for the deposition of metallic thin films. Dr. Pavel Leksin and Dr. Stephan Fuchs for the fruitful discussions about the construction of the ESR measuring setups. The support in the development of the experimental setups from the research technology department of the IFW Dresden and the clean room team headed by R. Engelhard (IFW Dresden) is greatly appreciated. This work is financed in part via the DFG research grant "Aktive verlustarme Magnetlager hoher Steifigkeit und Präzision mit integrierter Induktionsmessung und schneller Leistungselektronik" Nr. 221322256.

## Supplementary information:
## "Rolled-up self-assembly of compact magnetic inductors, transformers and resonators".

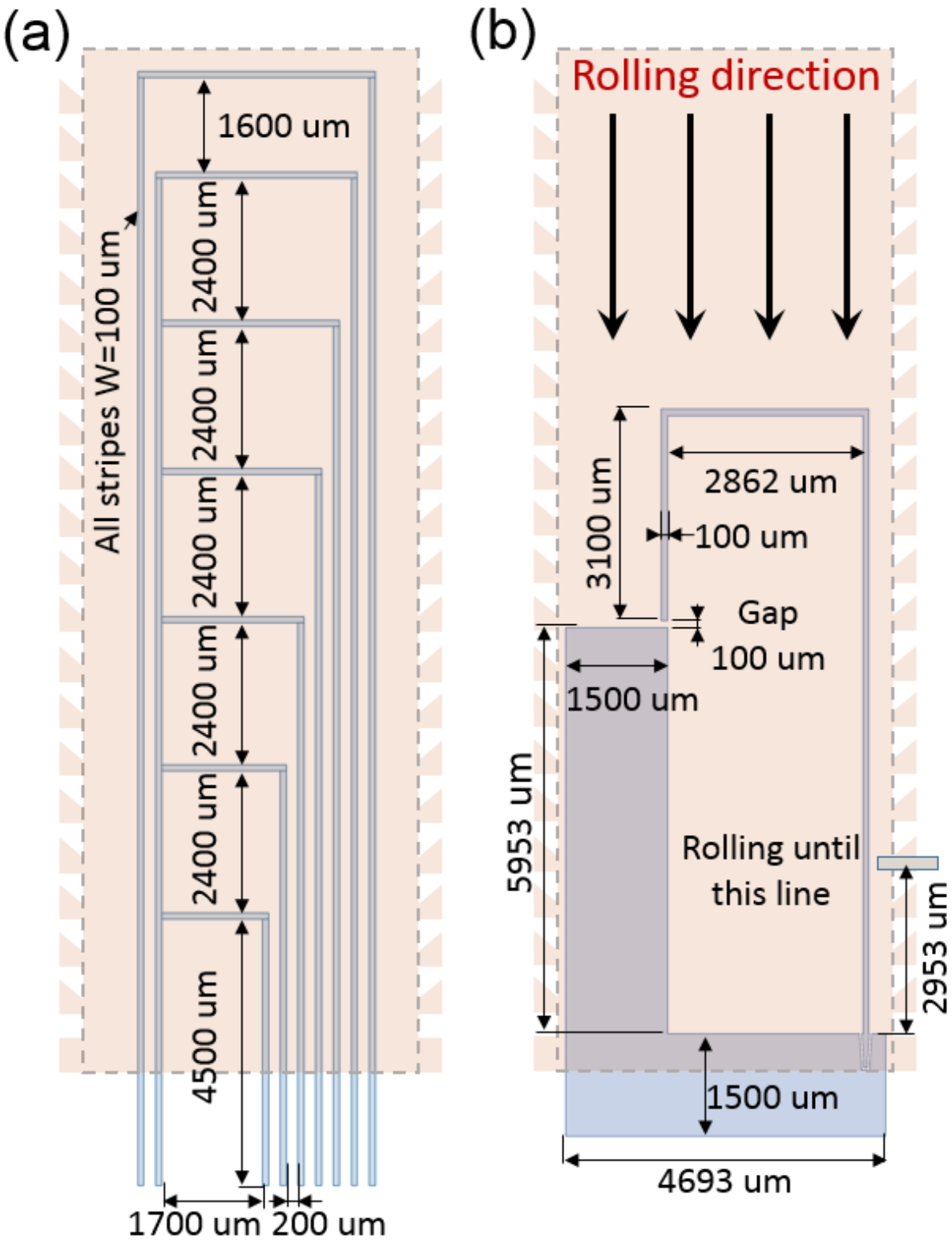

**Supplementary Figure 1.** Planar layouts applied for fabrication of self-assembled high quality micro coils and resonators (**a**) containing vertical stripes that were rolled into a set of axial coils and (**b**) an LC resonator structure. Horizontal lines connect left and right vertical stripes.

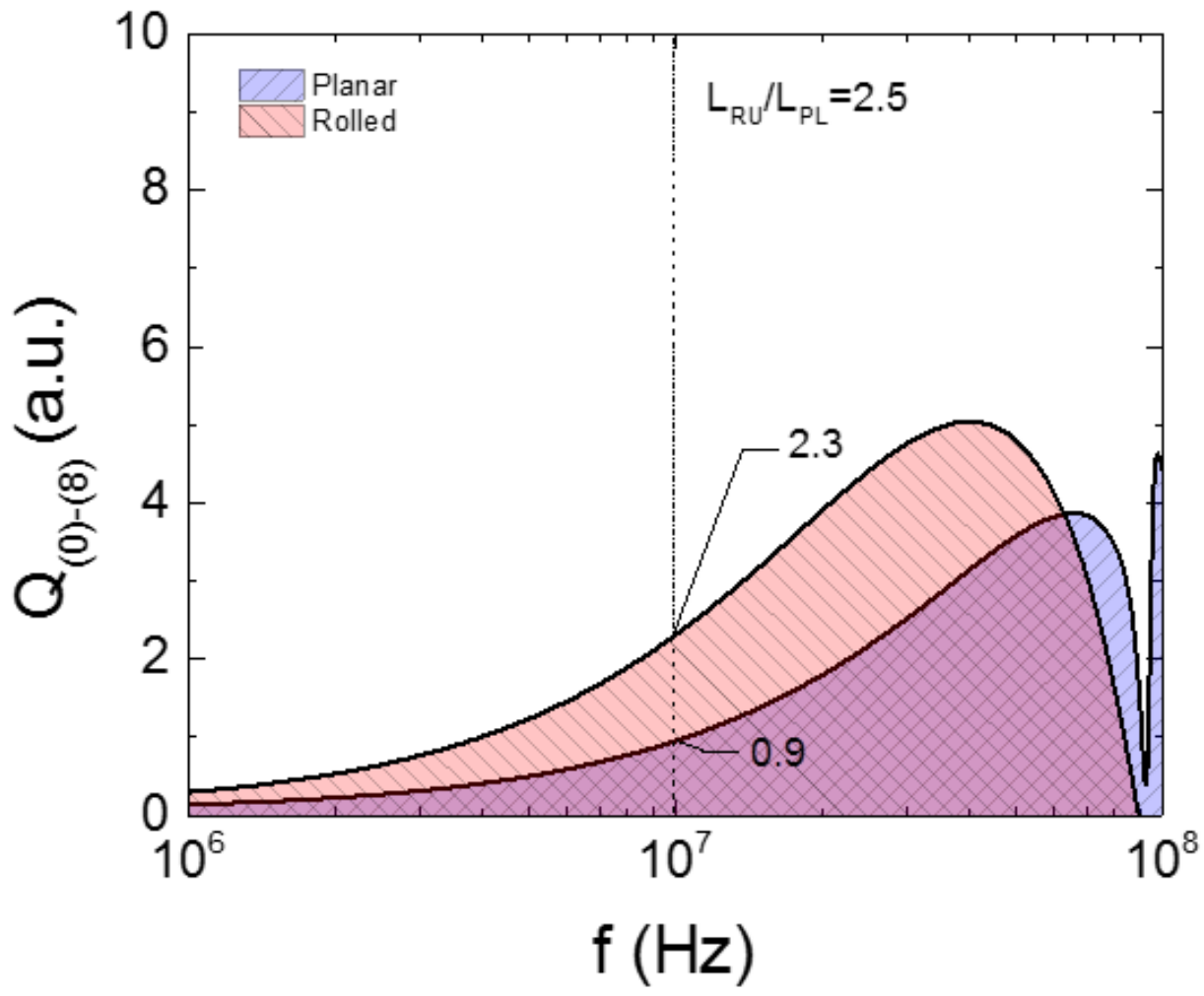

**Supplementary Figure 2.** Measured data represents the quality factor spectra of a coil set formed between electrodes (0) and (8) within a planar and self-assembled rolled up architectures. Particular values and their ratios are given at 10 MHz.

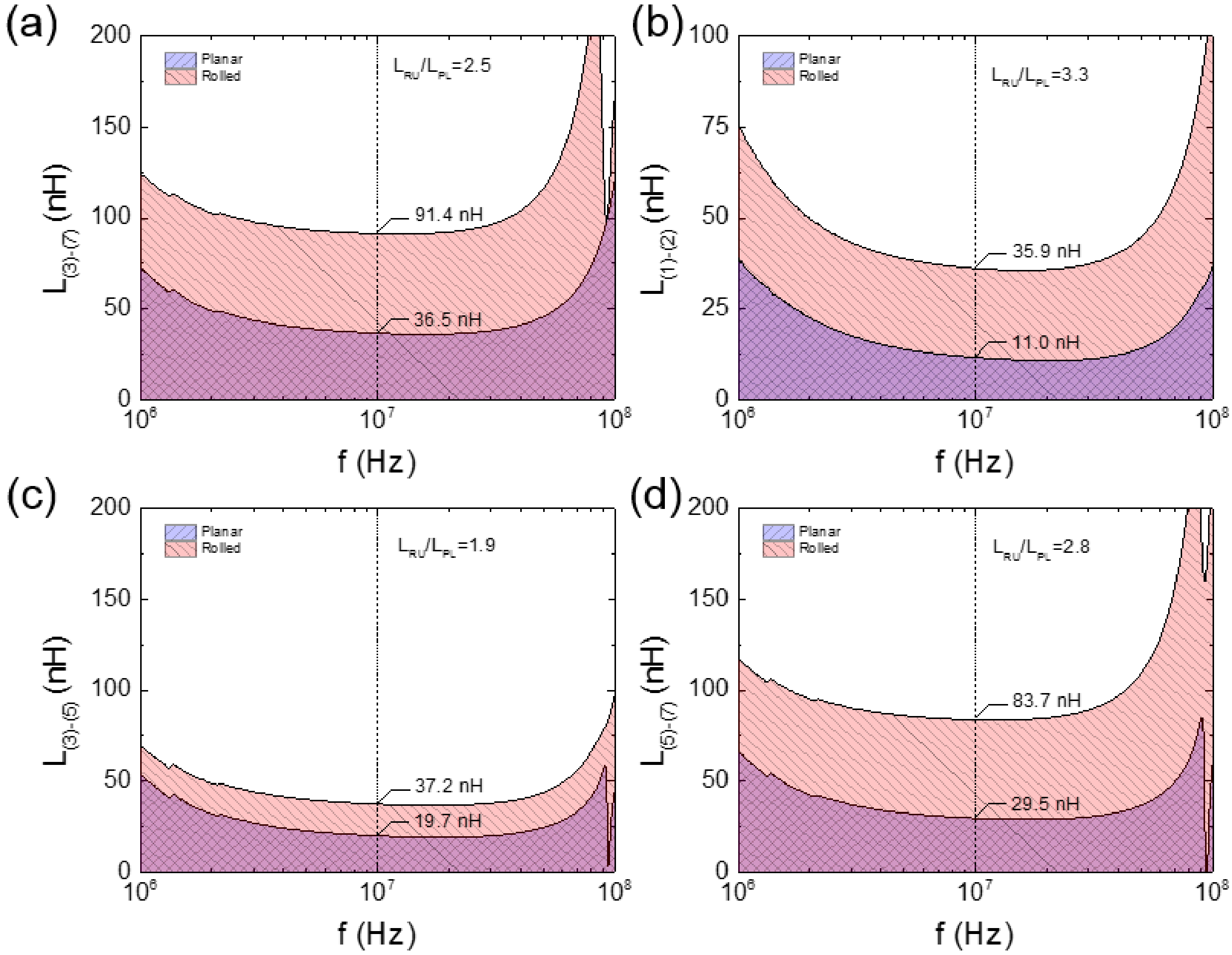

**Supplementary Figure 3.** Measured data represents the self-inductance spectra of different coil sets within a planar and self-assembled rolled up architectures. Particular values and their ratios

are given at 10 MHz. (**a**) The self-inductance of coils formed between electrodes (3) and (7). (**b**) The self-inductance of coils formed between electrodes (1) and (2). (**c**) The self-inductance of coils formed between electrodes (5) and (7). (**d**) The self-inductance of coils formed between electrodes (3) and (5).

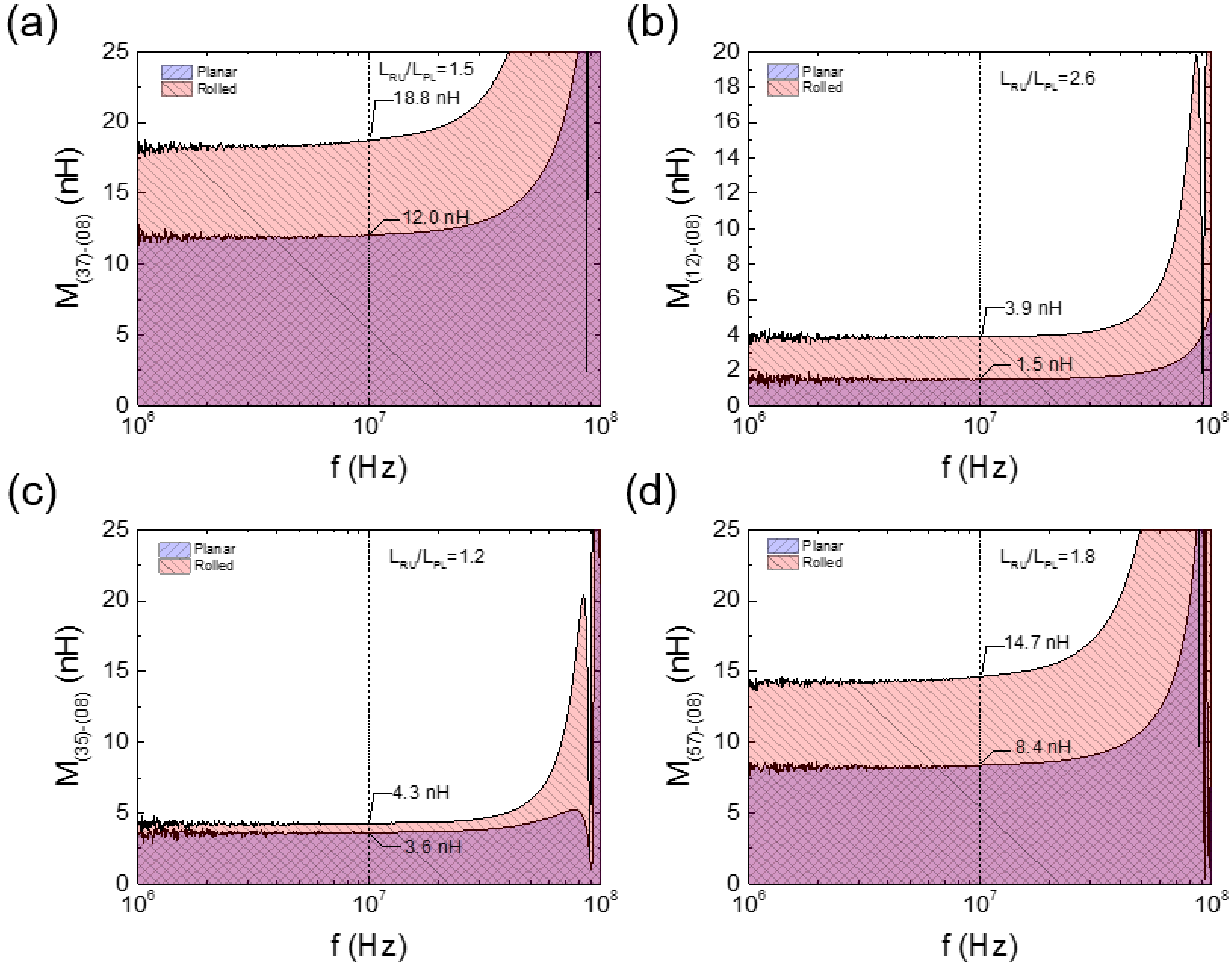

**Supplementary Figure 4.** Measured data represents the mutual inductance spectra among different coil sets within a planar and self-assembled rolled up architecture. Particular values and their ratios are given at 10 MHz. (**a**) The mutual inductance of coils formed between electrodes (37) and (08). (**b**) The mutual inductance of coils formed between electrodes (12) and (08). (**c**) The mutual inductance of coils formed between electrodes (57) and (08). (**d**) The mutual inductance of coils formed between electrodes (35) and (08).

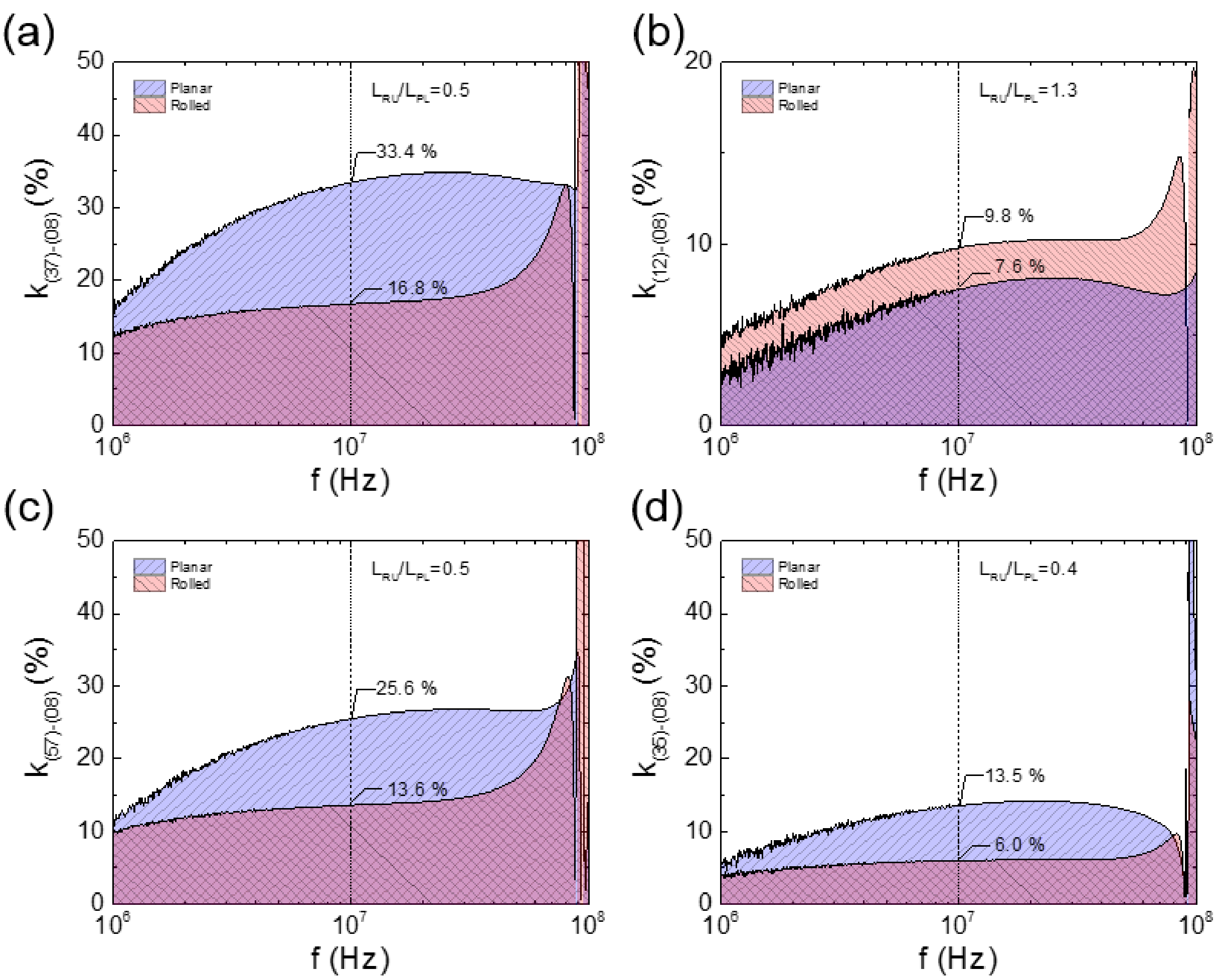

**Supplementary Figure 5.** Measured data represents the coupling coefficient spectra among different coil sets within a planar and self-assembled rolled up architecture. Particular values and their ratios are given at 10 MHz. (**a**) The coupling coefficient of coils formed between electrodes (37) and (08). (**b**) The coupling coefficient of coils formed between electrodes (12) and (08). (**c**) The coupling coefficient of coils formed between electrodes (57) and (08). (**d**) The coupling coefficient of coils formed between electrodes (35) and (08).

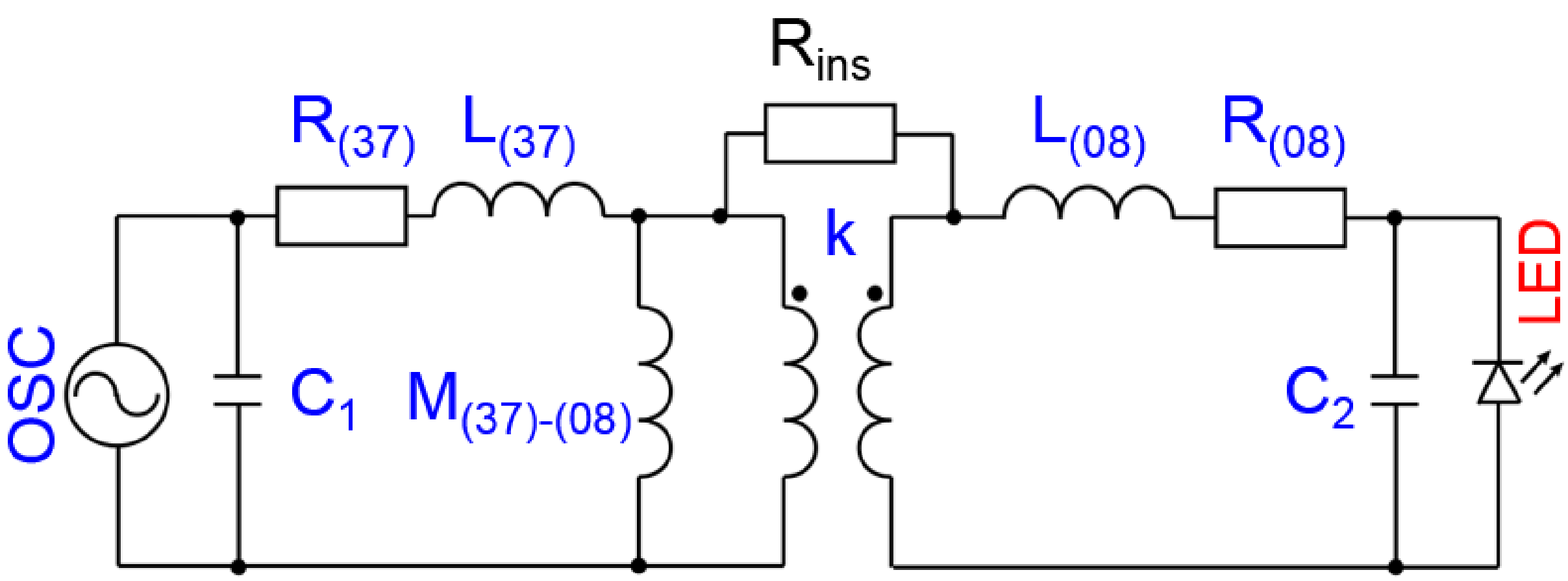

**Supplementary Figure 6.** Schematic that shows electrical connection between power generator transformer and the LED. The LED could be powered on the secondary side of the rolled-up transformer structure being DC insulated from the primary side of the transformer.